# Mechanical performance comparison of two surgical constructs for wrist four-corner arthrodesis via dorsal and radial approaches

Barthélémy Faudot [a,b,*], Julien Ballerini [c], Mark Ross [d], Philippe Bellemère [e], Benjamin Goislard de Monsabert [a], Laurent Vigouroux [a], Jean-Louis Milan [a,b]

[a] *Aix Marseille University, CNRS, ISM, Marseille, France*
[b] *APHM, Institute for Locomotion, Department of Orthopaedics and Traumatology, St Marguerite Hospital, Marseille, France*
[c] *NewClip Technics, Haute-Goulaine, France*
[d] *Brisbane Hand & UpperLimb, Brisbane, Australia*
[e] *Institut de la Main Nantes Atlantique, Nantes, France*

A B S T R A C T

*Background:* Four-corner arthrodesis, which involves fusing four carpal bones while removing the scaphoid bone, is a standard surgery for the treatment of advanced stages of wrist arthritis. Nowadays, it can be performed using a dorsal approach by fixing a plate to the bones and a new radial approach is in development. To date, there is no consensus on the biomechanically optimal and most reliable surgical construct for four-corner arthrodesis.
*Methods:* To evaluate them biomechanically and thus assist the surgeon in choosing the best implant orientation, radial or dorsal, the two different four-corner arthrodesis surgical constructs were virtually simulated on a 3D finite element model representing all major structures of the wrist. Two different realistic load sets were applied to the model, representing common tasks for the elderly.
*Findings:* Results consistency was assessed by comparing with the literature the force magnitude computed on the carpal bones. The Von Mises stress distribution in the radial and dorsal plates were calculated. Stress concentration was located at the plate-screw interface for both surgical constructs, with a maximum stress value of 413 MPa for the dorsal plate compared to 326 MPa for the radial plate, meaning that the stress levels are more unfavourable in the dorsal approach.
*Interpretation:* Although some bending stress was found in one load case, the radial plate was mechanically more robust in the other load case. Despite some limitations, this study provides, for the first time, quantified evidence that the newly developed radial surgical construct is mechanically as efficient as the dorsal surgical construct.

## 1. Introduction

Wrist arthritis results mainly from traumatic fractures in the intercarpal region or long-term conditions such as repetitive stress and carpal instability due to cartilage damage (Weiss and Rodner, 2007). The disease leads to abnormal joint kinematics, weakness, deformity and disabling pain (Watson and Ryu, 1986). The ability to perform simple everyday activities is compromised for patients since arthritis affects mobility, stability and strength, that are crucial for prehensile movements. The two most common patterns of wrist arthritis are scapholunate advanced collapse, occurring after attenuation of the scapholunate ligament, and scaphoid nonunion advanced collapse, occurring after a scaphoid fracture (Watson and Ballet, 1984; Watson and Ryu, 1986).

In advanced stages of the disease, surgery is currently used to help relieve symptoms (Shah and Stern, 2013). Four-corner arthrodesis (FCA) is a well-established surgery frequently used to treat wrist deformities like scaphoid nonunion advanced collapse and scapholunate advanced collapse (Bain and Watts, 2010; Shin, 2001; Strauch, 2011; Watson et al., 1981). The goal of FCA is to re-establish a stable and painless wrist, while both maintaining carpal alignment in an unimpinged flexion and preserving the maximum possible post-operative wrist motion and grip strength (Bain and Watts, 2010; Dacho et al., 2008; Watson et al., 1981). The lunate, capitate, hamate and triquetral bones are fused using a variety of fixation methods such as staples, screws, K-wires or plates (Shin, 2001) and the scaphoid bone, responsible for radioscaphoid arthritis, is partially or completely removed. The fusion eliminates motion between the carpal rows, leaving the motion to occur between the radius and a single rigid block of carpal bones. While some studies have reported favourable FCA outcomes for surgeries using, for example, K-wires (Ashmead et al., 1994; Watson et al., 1999) or plates (Merrell et al., 2008), others have found unacceptably high rates of non-union and major complications such as radio-carpal impingement and screw failure for surgeries using, for example, staples (Pauchard et al., 2014) or plates (Chung et al., 2006; Kendall et al., 2005; Shindle et al., 2007). To date, there is no consensus on the optimal and most reliable FCA surgical construct. Currently, the most widely applied is a dorsal approach, i.e. with a dorsal implant orientation, using a spider plate with the implant positioned centrally on the dorsal surfaces of the four bones. Two screws are inserted into each bone, except for the hamate bone, which has only one screw. A new radial plating approach is currently in development for FCA surgery to overcome the limitations of the dorsal approach mentioned above and to improve the surgery. The implant is positioned on the radial side of the lunate and the capitate with five screws, each crossing two bones. The plate is not flat but has an offset, so as to follow the geometry of the lunate (Viegas et al., 1990).

It is not well understood how complications can occur (e.g. loosening), as it is difficult to directly measure the internal loadings of the wrist. Numerical models allow the estimation of internal loads that need to be validated by comparison with in-vivo data. The numerical data supplement the observations performed at the functional level in traditional clinical trials. Over the past decade, researchers have developed 3D finite element (FE) models of the wrist, both for the intact wrist (Gislason et al., 2009) and for the arthritic wrist (Bajuri et al., 2012). Yet in terms of numerical simulations of wrist surgery, only a few studies have focused on modelling wrist arthrodesis. In (Gislason et al., 2012; Iwasaki et al., 1998; Majors and Wayne, 2011), different wrist arthrodesis surgeries were modelled and compared with each other as well as with a healthy wrist, in terms of range of motion, joint contact forces and joint contact pressure. However, these three studies did not model the FCA arthrodesis and just considered the fused bones without modelling the surgical construct. In (Bicen et al., 2015), (Dvinskikh et al., 2011) and (Márquez-Florez et al., 2016), although the FCA surgery was evaluated in terms of joint load distribution or wrist range of motion, the former used a two-dimensional model, and none modelled the implant and screws to investigate pressure distribution and intensity in the surgical construct. To date, there has been no in silico evaluation of different FCA surgical approach, in particular the radial surgical approach, representing the surgical construct used for the treatment of wrist arthritis.

The aim of this study was therefore to develop a 3D numerical model of the wrist to evaluate the mechanical performance of radial and dorsal FCA surgical approaches through virtual surgery. The FE model of the wrist was composed of bones, cartilage, ligaments and each of the two radial and dorsal surgical constructs. These two FCA surgical constructs were evaluated and compared in terms of mechanical stresses in the surgical construct and at the bone-implant interface.

## 2. Methods

### 2.1. Healthy model of the wrist

#### 2.1.1. CT acquisition

Computed tomography (CT) images of the hand in a neutral position of a male subject were acquired (age: 35 years). The subject had a fracture of the distal radius and signed an informed consent form. The protocol was approved by the local ethics committee. The CT system was a Toshiba-MEC CT3 (GE Medical Systems; Chicago, USA) (150 mA × 120 kV; slice thickness 625 μm).

#### 2.1.2. Modelling bone and cartilage

Bone segmentation was performed after the CT-scan acquisition using the 3D image reconstruction software Mimics 22.0 (Materialise; Leuven, Belgium). Bones from the distal radius to the proximal end of the metacarpal bones were meshed using quadratic tetrahedral elements (C3D10) with a maximum element edge length of 1.8 mm, determined after mesh convergence analysis. The solid geometries were imported into the finite element software Abaqus (Dassault Systemes; Vélizy-Villacoublay, France). Bones were assumed to be composed of an inner cancellous tissue with a 2.6 mm outer cortical layer (Louis et al., 1995). Cortical (E = 18 GPa, ν = 0.2) and cancellous bone (E = 300 MPa, ν = 0.25) were modelled as linear elastic isotropic materials, see Table 1. Cartilage was created by identifying joint surfaces on the bones and then extruding surface elements to create tetrahedral elements. Cartilage thickness varied across the joints with a thickness of 0.8 mm and 0.7 mm in average at the radiocarpal and intercarpal joints, respectively, which is consistent with literature data (Moore et al., 2011). Cartilage was modelled using a Neo-Hookean hyper-elastic material detailed in Table 1. Neo-Hookean constants ($C_{10}$ and $D_1$) were calculated, as in Eq. (1), based on a Young's modulus of 1.64 MPa and a Poisson's ratio of 0.2 (Dourthe et al., 2019).

$$C_{10} = \frac{E}{4(1+\nu)} \text{ and } D_1 = \frac{6(1-2\nu)}{E} \quad (1)$$

The number of elements for the whole structure was 352,050 with an element density of 6.4 elements/mm$^3$.

#### 2.1.3. Modelling ligaments and TFCC

Extrinsic and intrinsic carpal ligaments were modelled based on anatomical studies (Berger, 2001; Kijima and Viegas, 2009). Distributed insertions were simulated by applying non-linear spring elements (CONN3D2) in parallel with tension-only behaviour. Ligament stiffness was derived from (Eschweiler et al., 2016; Majors and Wayne, 2011). Carpometacarpal and radiocollateral ligaments were modelled as rigid connectors to maintain stability, facilitate convergence and work against the input forces of the first metacarpal bone, as the model did not include all tendons, muscles and soft tissues constraining the carpus. The triangular fibrocartilage complex (TFCC), assumed to be a support for the triquetral and lunate bones (Palmer and Werner, 1981), was modelled as multiple springs connecting the ulna with the carpus. Spring stiffness values representing the TFCC were similar to those of the other articular cartilage of the wrist, which consists of fibrocartilage tissue bearing compressive loads (Anderson et al., 2005; Dvinskikh et al., 2011).

#### 2.1.4. Boundary conditions

Bone and cartilage were fixed together to prevent relative motion. Interaction contact was modelled at the radiocarpal and proximal intercarpal joints with a friction coefficient of 0.02 (Wright and Dowson, 1976). A tie contact was applied to the carpometacarpal joints and the trapezium-trapezoid joint, both showing little movement (Kauer, 1986).

### 2.2. FCA virtual surgery

The circular spider implant of the dorsal construct was scanned using an industrial scanner (Artec Space Spider; Artec 3D, Luxembourg) to acquire the external geometry of the implant. FCA with scaphoid excision was modelled by positioning the plate and the screws in a way that represented each of the two surgical constructs. The lunate bone was positioned in a neutral alignment in the sagittal plane with respect to the capitate according to the surgical technique (Watson et al., 1999). Cartilage structures between the four bones were removed (Enna et al., 2005). The radial construct was positioned with the offset plate and five locking screws and the dorsal construct was positioned with the circular spider plate and seven locking screws. The screws used in this study were of different lengths while the plate was designed so that the screws can be placed with variable angulation to match the specific geometry of each carpal bone involved in the carpal fusion. Under the supervision of two orthopaedic surgeons specialised in hand and upper limb surgery, particular care was taken to correctly position the plate and screws. Radioscaphocapitate and long radiolunate ligaments were kept, preventing ulnar translocation of the remaining carpal bones (Shin, 2001). The screws were bonded to the bones, with the screw threads omitted for numerical simplification. Contact on the bone-plate surfaces was modelled using a penalty-based contact condition with a normal contact stiffness of 600 N.mm$^{-1}$ (Bernakiewicz and Viceconti, 2002) and a friction coefficient of 0.37 (Hayes and Perren, 1972).

**Table 1**
Material properties and element types of the wrist finite element model.

| Component | Element type | Constitutive model | Constants |
|---|---|---|---|
| Cortical bone | Tetrahedral | Linear elastic | E = 18 GPa; ν = 0.2 |
| Cancellous bone | Tetrahedral | Linear elastic | E = 300 MPa; ν = 0.25 |
| Cartilage | Wedge | Hyper-elastic Neo-Hookean | $C_{10}$ = 0.28 MPa; $D_1$ = 0.15 MPa$^{-1}$ |
| Ligaments | Tension-only connector | Non-linear elastic | From 40 N/mm to 150 N/mm |

To improve the stability of the numerical model by constraining the displacements of the trapezium and trapezoid bones, multiple compression springs were modelled at the site of the excised scaphoid bone. These springs were inserted proximally on the radius and distally on the trapezoid and trapezium bones, with total stiffness assumed to be the same as that of the healthy cartilage structures (Dvinskikh et al., 2011).

The plate and screws were composed of grade 4 titanium *Ti Grade 4* and titanium alloy *TiAl$_6$V$_4$*, respectively. Material properties were assumed to be linear elastic isotropic with a Young's modulus of 105 GPa and 110 GPa and a Poisson's coefficient of 0.33 and 0.33 for the plate and the screws, respectively. The material properties of the surgical constructs are given in Table 2.

### 2.3. Loads applied to the model

Two different sets of physiologically relevant loads were applied to the model. These loads were chosen because these tasks are commonly used in everyday life (Napier, 1956).

First, a set of realistic loads was applied to the five metacarpal bones along their primary axis, to represent a maximum strength gripping task (Fig. 1). This task is considered as critical and represents the upper limits of the physiological loads (Gislason et al., 2009). Loads were halved to represent the loss of postoperative strength compared to the opposite hand (Kendall et al., 2005), and to remain in the range of everyday task intensity of the elderly population. Forces were previously calculated in (Fowler and Nicol, 2001) from an inverse dynamic biomechanical model using experimentally measured external forces as input. The magnitude of the loads applied to each metacarpal bone is shown in Table 3. The proximal end of the radius and ulna were fully constrained, and the metacarpal bones were constrained along the direction of their primary axis.

Second, maximum strength in a hyperextended position was modelled to represent a common task for the elderly, i.e. lifting themselves out of a chair using their wrists (Alexander et al., 2000). Maximum wrist extension was measured at an average of 35° postoperatively (Chung et al., 2006; Merrell et al., 2008). Thus, the wrist was moved to a hyperextended position following (Moojen et al., 2002) with an extension of 30° relative to the radius. This set of loads corresponded to the contact forces of the palm of the hand when pressing a flat surface at maximum voluntary contraction (Figueroa-Jacinto et al., 2018). Loads were evenly distributed on the lunate and triquetral bones with the magnitudes shown in Table 4. The proximal end of the radius and ulna were fully constrained (Fig. 2).

### 2.4. Numerical model outputs and assessment

The analyses were carried out using the Abaqus explicit solver run over a simulation period of 0.01 s with an average CPU time to solve each simulation of 20–24 h. The ratio of forces transmitted to the radius and ulna during the load sets were calculated for the healthy model. Force transmitted to the radiocarpal joint was calculated for both the healthy and the arthrodesis wrist by summing the contact force on all cartilage surfaces and ligaments. These numerical values were compared with in-vivo and in-vitro results of the literature to assess the numerical validity of the model with an acceptable agreement threshold which we set to ten percentage points. The Von Mises stress distribution in the screws and plate and at the implant-bone interface was calculated for both radial and dorsal plate models.

## 3. Results

### 3.1. Numerical model assessment

The force transmission ratio between the radius and ulna for the healthy wrist was 86% transmitted to the radius and 14% to the ulna. The force transmission ratio at the radioscapholunate joint was 66% transmitted to the radioscaphoid joint and 34% transmitted to the radiolunate joint. After the FCA surgery and scaphoid excision, relative loads at the radiocarpal joint changed radically. The force transmission ratio after the scaphoidectomy carpal fusion was 9% transmitted to the radioscaphoid joint via the compression springs, 74% to the radiolunate joint and 17% to the TFCC.

The kinetic energy of the numerical model did not exceed 5% of the strain energy, excluding any dynamic behaviour. This FE model can therefore be considered as a quasi-static model (Harewood and McHugh, 2007).

### 3.2. Surgical construct mechanical stress

Von Mises stress distribution was calculated for both load sets on the radial and dorsal surgical constructs. Only 2% of the elements exceeded values of 500 MPa for the radial plate and the dorsal plate construct. These elements were at the plate-screw interface on the external surfaces due to the bonded interface, which fused the two surfaces without any micro-movement being allowed. This condition of perfect fusion led to localised stress and strain which were discarded as numerical artefacts. The plate-screw interface is defined as the hole areas of the plate in contact with the screw heads.

#### 3.2.1. Maximum strength gripping task

Stress concentration was located at the plate-screw interface for both radial and dorsal surgical constructs, with a peak stress value of 352 MPa for the dorsal plate and 318 MPa for the radial plate (Fig. 3). A large stress zone with an average magnitude of 171 MPa was found in the radial plate at the central part of the plate on the offset. Bending stress was found in the two longest screws of the radial surgical construct at the interface between the two bones.

#### 3.2.2. Maximum strength in a hyperextended position

For the hyperextended position, stress concentration was higher in the dorsal surgical construct than in the radial surgical construct (Fig. 4). Stress concentration was located at the plate-screw interface in both surgical constructs. This load case produced higher stress than the maximum strength gripping task load, with peak stress of 414 MPa for the dorsal plate.

**Table 2**
Material properties and element types of the plate and screws of radial and dorsal constructs.

| Component | Element type | Constitutive model | Constants |
|---|---|---|---|
| Plate | Tetrahedral | Linear elastic | E = 105 GPa; ν = 0.33 |
| Screw | Tetrahedral | Linear elastic | E = 110 GPa; ν = 0.33 |

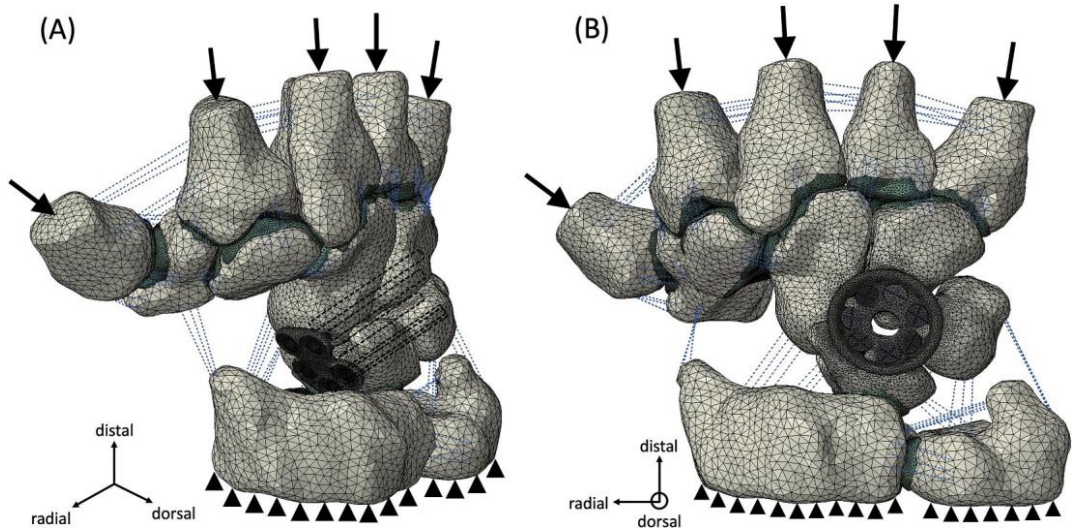

**Fig. 1.** Finite element model of the wrist in a neutral position including bones, cartilage, and ligaments and representing a maximum strength gripping task. The four-corner arthrodesis with scaphoidectomy was modelled to represent the two surgical approaches with (A) the radial surgical construct involving a radial plate with five screws and (B) the dorsal surgical construct involving a dorsal plate with seven screws. Cartilage in green was modelled by extrusion of the bone surfaces, ligaments in blue were represented by multiple non-linear spring elements. Loads were applied on metacarpal bones along their primary axis and the radius and ulna were fixed. (For interpretation of the references to colour in this figure legend, the reader is referred to the web version of this article.)

**Table 3**
Loads applied to the metacarpal bones along their primary axis during the maximum strength gripping task.

| Metacarpal bone | First | Second | Third | Fourth | Fifth |
| --- | --- | --- | --- | --- | --- |
| Loads | 112.8 N | 60.2 N | 53.2 N | 44.0 N | 38.7 N |

**Table 4**
Loads applied to the carpal bones in the dorsal direction during the maximum strength in a hyperextended position.

| Carpal bone | Lunate | Triquetral |
| --- | --- | --- |
| Loads | 100 N | 100 N |

## 4. Discussion

This study is the first to investigate the mechanical performance of a new method for the treatment of wrist osteoarthritis by FCA of the carpal bones using a radial implant orientation. For this purpose, a three-dimensional FE model of the wrist after FCA virtual surgery was developed, seeking to assess and compare stresses of radial and dorsal surgical constructs during two common tasks in the elderly. The mechanical behaviour of wrist surgical constructs remains poorly understood, with only a few studies focused on total wrist arthroplasty (Bajuri et al., 2013; Gislason et al., 2017; Grosland et al., 2004). As far as the authors know, this is the first FE model of an arthrodesis wrist representing the implant with all major wrist structures and driven by realistic load sets.

Comparisons with data from the literature were carried out to assess the validity of the model. Calculated relative loads on the radius and ulna were consistent with cadaver studies of healthy wrists (Palmer and Werner, 1984; Werner et al., 1992) which showed a force transmission ratio of 80% to the radius and 20% to the ulna, resulting in a difference of only six percentage points with the model. Similarly, the relative load transmitted to the radiocarpal joint was consistent with the literature (Hara et al., 1992; Skie et al., 2007), with a force transmission ratio of 50% to the scaphoid, 35% to the lunate and 15% to the TFCC. Scaphoid fossa contact forces after scaphoidectomy were negligible compared to radiolunate joint contact force, which is in good agreement with the literature showing a statistically significant decrease in scaphoid fossa loading (Dvinskikh et al., 2011; Skie et al., 2007) after FCA with scaphoid excision. This showed that the totality of the forces transmitted in the carpus will be carried by the implanted device making the plate and screws mechanically loaded. Regarding the loads applied to the model, intrinsic muscles act as dynamic stabilisers of the wrist. However, they were not included in the numerical model as in other FE studies for reasons of simplification (Gislason et al., 2017). Instead, we limited wrist movement by constraining the metacarpal bones to move in their primary axis and by modelling no relative movement in the carpometacarpal joints. The transmission of forces in this numerical model is therefore physiological for both the healthy and the arthrodesis wrist model. Other manual tasks and surgical constructs can be investigated in future studies.

The two FCA surgical constructs can therefore be evaluated and compared in terms of mechanical stresses in the implant and at the bone-implant interface. The overall stress in both surgical constructs is high, with peak stress at 414 MPa and large stress zones at 150–200 MPa. These high stresses are mainly due to the loading conditions, which are considered the worst-case scenario, and the non-representation of the bone graft between the four fused bones, which can unload the surgical constructs. Under the load set in the maximum strength grip task, the highest stress was located at the plate-screw interface for both surgical constructs. This is consistent with reported clinical findings of a high number of screw breaks at the head-shrink interface for the dorsal plate (Shindle et al., 2007). The stress in the screws was higher in the hamate screw for the dorsal surgical construct as there is only one screw for this bone versus two for the other three carpal bones. The stress distribution in the radial plate is more homogeneous than in the dorsal surgical construct, with a large stress zone in the radial plate located in the centre of the plate providing better mechanical strength. Bending stress in the two longest screws of the radial surgical construct could lead to the possibility of lower mechanical performance of the radial plate compared to the dorsal plate, as the screws could break at mid-length under fatigue. However, Von Mises stress levels in this area are far from the yield strength of titanium are therefore not critical. Under the load set for maximum strength in a hyperextended position, stress concentration was higher in the dorsal plate than in the radial plate and located at the plate-screw interface for both approaches. This load case provided the highest stress (peak stress at 414 MPa) and appears to be the most critical load, in support of clinical findings (DeGoede and Ashton-Miller, 2002; Dunn, 1972) of a high prevalence of wrist fractures, in particular of the scaphoid, during falls. Depending on the tasks and activities performed by the patient, e.g. grasping tasks or tasks requiring pushing with the hand, the surgical construct will not be loaded mechanically in the same way. However, the surgical approach may not be guided by the type of activities the patient is accustomed to performing, as each of the two tasks described in this study is used by patients on a daily basis. Thus, taking into account the two loading cases, the radial plate, in development, is mechanically as efficient as the currently developed dorsal plate.

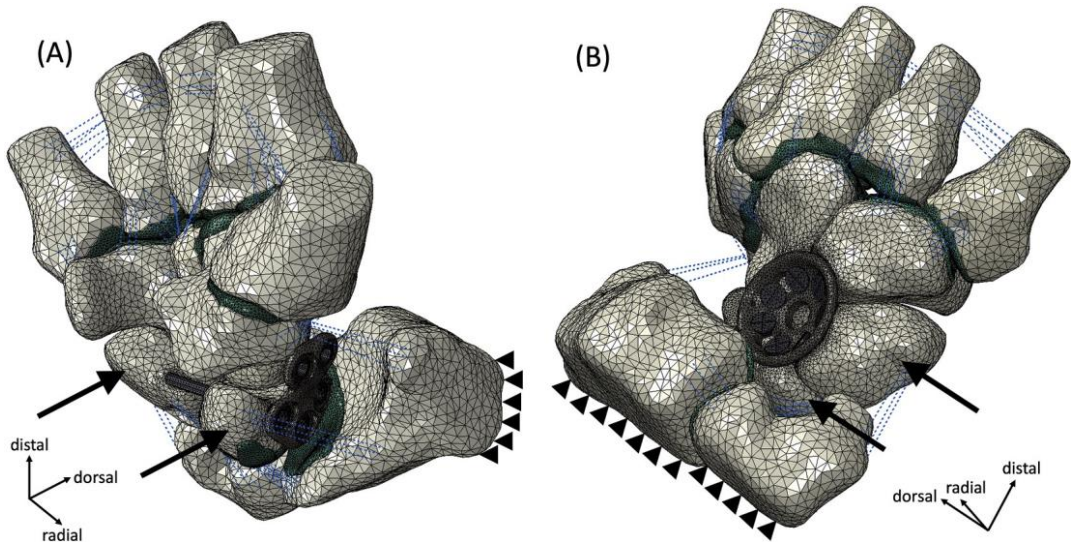

**Fig. 2.** Finite element model of the wrist in a hyperextended position including bones, cartilage, and ligaments and representing a common task of lifting from a chair using the wrists. The four-corner arthrodesis with scaphoidectomy was modelled to represent the two surgical approaches with (A) the radial surgical construct involving a radial plate with five screws and (B) the dorsal surgical construct involving a dorsal plate with seven screws. Cartilage in green was modelled by extrusion of the bone surfaces, ligaments in blue were represented by multiple non-linear spring elements. Loads were applied to the lunate and triquetral bones in the dorsal direction and the radius and ulna were fixed. (For interpretation of the references to colour in this figure legend, the reader is referred to the web version of this article.)

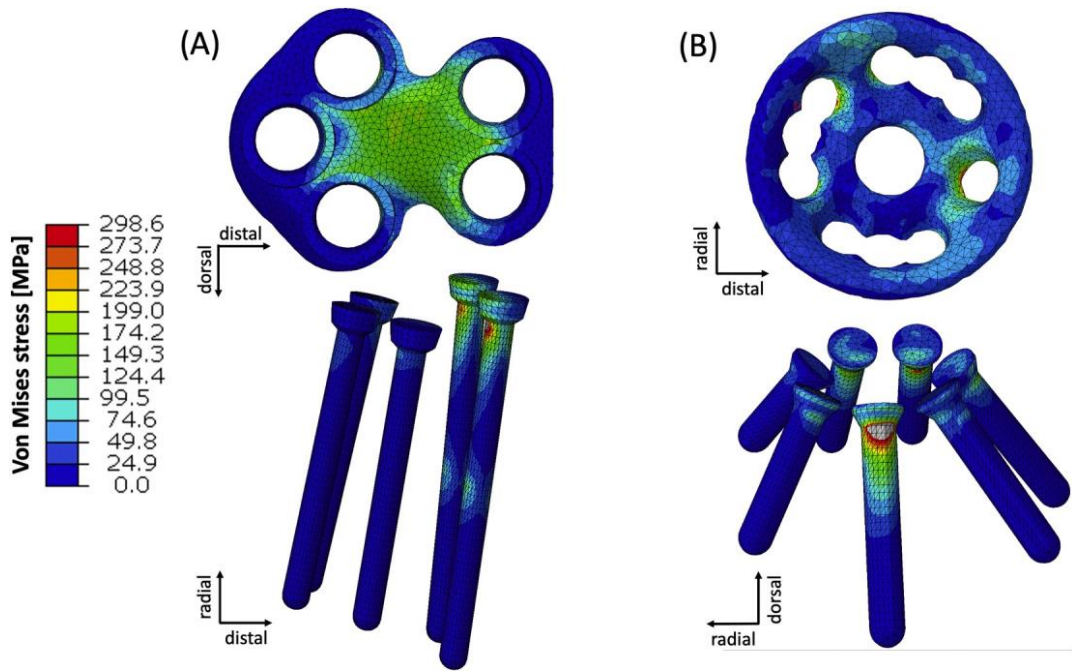

**Fig. 3.** Von Mises stress distribution in the plate and screws of the radial (A) and dorsal (B) surgical constructs during a maximum strength gripping task with the wrist in a neutral position.

The radial surgical construct is therefore mechanically promising according to the simulations and has major differences in surgical technique compared to the dorsal surgical construct. The surgical technique for the dorsal plate requires a reaming on the dorsal surface of the four bones to allow the insertion of the spider plate without protruding from the dorsal surface of the bones. On the contrary, the cortical bone is not removed for the radial plate, resulting in less stress at the bone-implant interface. Furthermore, the impact of trabecular bone stiffness on Von Mises stress in both surgical constructs was evaluated by varying the Young's modulus of the trabecular bone. By varying the Young's modulus of the trabecular bone by $+/-10\%$ from the initial value of 300 MPa, considered as reference, the distribution of the Von Mises stress was equivalent to the reference simulation. Moreover, varying the trabecular bone stiffness does not discriminate one surgical construct over the other regarding the Von Mises stress distribution. The radial plate construct is more versatile than the dorsal one because it can be used both for the arthrodesis of four bones as described in this article and also for the arthrodesis of the only two radial bones alone, capitate and lunate bones. Another advantage of the FCA surgical technique using the radial plate compared to the dorsal plate is that, during the insertion of the plate and screws, the construct is not in conflict with the compression device that holds the four bones in the correct position. Finally, the radial plate construct is anatomically less invasive than the dorsal plate construct with less devascularisation, less invasive with respect to the extensor tendons and allows a better wrist range of motion with less impingement between the plate and the radius during hyperextension of the wrist.

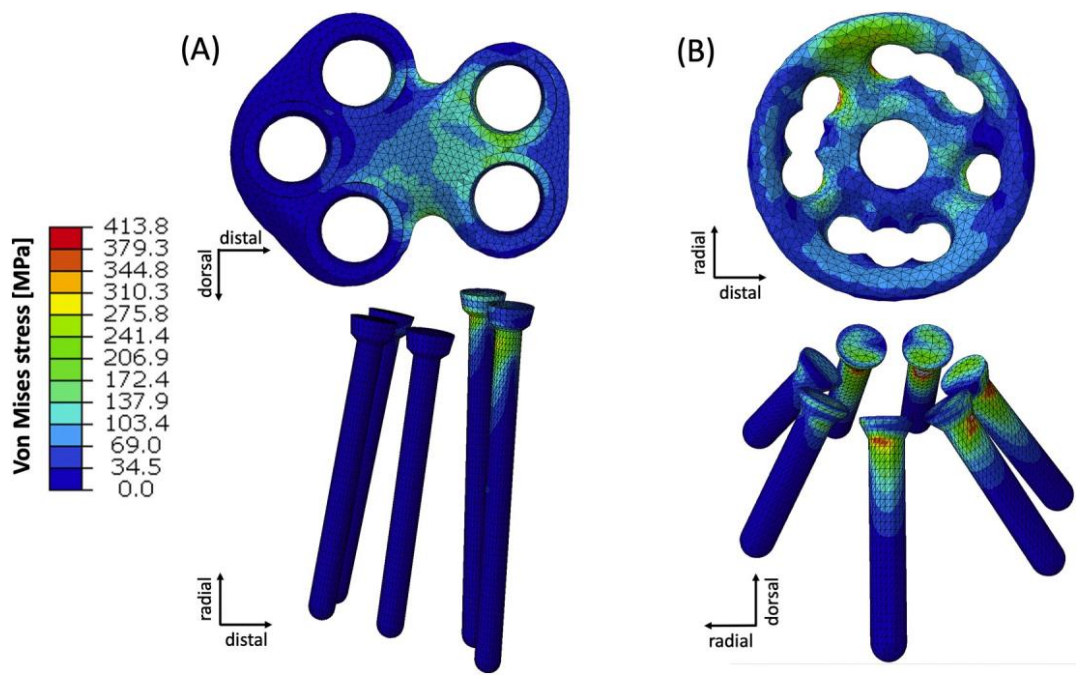

**Fig. 4.** Von Mises stress distribution in the plate and screws of the radial (A) and dorsal (B) surgical constructs during a common task of lifting from a chair using the wrists with the wrist in a hyperextended position.

Despite the significant results demonstrated in this study, this numerical model is subject to limitations. Firstly, screw threads were not represented in this study because, as demonstrated (Inzana et al., 2016), the bonded interface with screw simplification serves as an effective method of comparing implants using the same thread profile. However, this simplification makes it impossible to compare the screw pull-out between the two constructs. The dorsal plate is known to have a high non-union rate (Chung et al., 2006; Kendall et al., 2005; Shindle et al., 2007). It would therefore be interesting to compare the pull-out force between the two surgical constructs to estimate implant stability since pull-out force is related to implant loosening. Secondly, this study focused on a single bone geometry in order to compare the mechanical stresses between the two surgical constructs. Radial and dorsal plates were therefore compared for a single patient subjected to two extreme tasks of everyday life. Creating FE wrist models is time-consuming, ruling out a large cohort of models, which is why anatomical differences and age-related differences in bone properties were not considered. A larger study including several subjects with anthropometric variations and different bone qualities would enable different plate placements and screw lengths to be explored. The two surgical constructs could then be compared on a wider variety of patients and thus reveal significant differences between the two approaches.

This study is the first to provide quantitative data regarding the biomechanical suitability of the radial plate approach as an alternative to the dorsal plate approach for four-corner arthrodesis with scaphoidectomy. For this purpose, a virtual surgery was performed on a 3D finite element model of the wrist. Even though the radial plate has its limitations, i.e. screw bending when grasping, it significantly reduces mechanical stress during the common task of lifting out of a chair. Both surgical constructs have high-stress concentrations in the screws at the head-shrink interface. The results suggest that the new radial surgical construct currently in development is mechanically promising. Clinical studies involving functional tests should also be performed to compare the two surgical constructs and assess the long-term viability of the new construct, as clinical reality can be more complex than in silico simulated situations.

### Declaration of Competing Interest

The authors whose names are listed below certify that they have no affiliations with or involvement in any organization or entity with any financial interest or non-financial interest in the subject matter or materials discussed in this article.

### Acknowledgments

This study is part of a research contract established between the company NewClip Technics and Aix-Marseille University. No conflict of interest exists between the authors.